\documentclass[aps,preprint,showpacs,showkeys,superscriptaddress,tightenlines]{revtex4}
\usepackage{graphicx}
\usepackage{amssymb}
\usepackage{bm}

\def\be{\begin{eqnarray}}
\def\ee{\end{eqnarray}}

\begin{document}

\bibliographystyle{prsty}



\title[]{Pentaquark $\Theta^+$ Mass and Width in Dense Matter}
\author{Hyun-Chul \surname{Kim}}
\email{hchkim@pusan.ac.kr}
\affiliation{Department of Physics and  Nuclear Physics \& Radiation
Technology Institute (NuRI), Pusan National University, 609-735
Busan, Republic of Korea}
\author{Chang-Hwan \surname{Lee}}
\email{clee@pusan.ac.kr}
\affiliation{Department of Physics and  Nuclear Physics \& Radiation
Technology Institute (NuRI), Pusan National University, 609-735
Busan, Republic of Korea}
\author{Hee-Jung \surname{Lee}}
\email{Heejung.Lee@vu.es}
\affiliation{Department of Physics and  Nuclear Physics \& Radiation
Technology Institute (NuRI), Pusan National University, 609-735
Busan, Republic of Korea}
\affiliation{Departament de F\'{i}sica Te\`{o}rica,
Universitat de Val\`{e}ncia, E-46100 Burjassot (Val\`{e}ncia), Spain}

\date[]{Received December 2004}

\begin{abstract}
We investigate medium modifications of the pentaquark $\Theta^+$
in dense medium, taking into account different parities of the exotic 
$\Theta^+$ baryon.  We find that the chemical potential of the
$\Theta^+$ is shifted in a density-dependent way to one-loop order.
We also investigated the effect of the scaled nucleon mass in dense
medium on the $\Theta^+$ propagator.  The results turn out to depend 
sensitively on the scaled nucleon mass and on the parity of the  
$\Theta^+$.   
\end{abstract}
\maketitle

\setcounter{footnote}{0}

Since the discovery of the pentaquark
($\mathrm{uudd}\bar{\mathrm{s}}$) baryon 
$\Theta^{+}$~\cite{Nakano:2003qx,Stepanyan:2003qr,Kubarovsky:2003fi,
Barmin:2003vv,Barth:2003es,Airapetian:2003ri}, which was
motivated by the work of Diakonov {\em et al.}~\cite{Diakonov:1997mm},
its properties~\cite{quark,Ohetal,chiral,chiral2,bound,qsr,lqcd}
and production mechanism~\cite{reaction} have been extensively
investigated.  The unique feature of the 
$\Theta^+$ lies in its small mass (1540 MeV) and very narrow width
($<25$ MeV), which is shared by the recently found pentaquark state
$\Xi_{3/2}$~\cite{Alt:2003vb}.  Since the $\Theta^+$ is known to decay
into a neutron and a $K^+$, its strangeness must be
$+1$~\cite{Nakano:2003qx}.  The isospin of the $\Theta^+$ is concluded
to be zero~\cite{Barth:2003es,Airapetian:2003ri}.  However, its spin
and parity have not been measured experimentally yet.  In spite of the great
amount of theoretical work on the $\Theta^+$, there is no agreement on
its spin and parity.  For example, chiral models advocate a positive
parity for the $\Theta^+$~\cite{chiral} whereas lattice quantum
chromodynamics (QCD) and
the QCD sum rule claim that its parity should be
negative~\cite{lqcd,qsr}.

Recent works on the $\Theta^+$ have concentrated on how to determine
its parity~\cite{Nakayama:2003by,Zhao:2003bm,Yu:2003eq,
Carlson:2003xb,Thomas:2003ak,Hanhart:2003xp,Nam:2004qy,Nam:2004hk,
Nam:2004ph,Mehen:2004dy,Rekalo:2004kb}.  In particular, Thomas {\em et
al.}~\cite{Thomas:2003ak} have put forward an
unambiguous method to determine the parity of the $\Theta^+$ via
polarized proton-proton scattering at and just above the threshold of the
$\Theta^+$ and $\Sigma^+$: If the parity of $\Theta^+$ is positive,
the reaction is allowed at the threshold region only when the spin of
two protons is $S = 0$. On the other hand, if the parity is
negative, the reaction is allowed
only when $S=1$.  
Hanhart {\em et al.}~\cite{Hanhart:2003xp} 
extended the work of Thomas {\em et al.}~\cite{Thomas:2003ak} 
to determine the parity
of the $\Theta^+$, asserting that the sign of the spin correlation
function $A_{xx}$ agrees with the parity of the $\Theta^+$ near
threshold.  Similarly, Rekalo and
Tomasi-Gustafsson~\cite{Rekalo:2004kb} proposed methods for
pinning down its parity via a measurement of the spin correlation
coefficients in three different reactions, {\em i.e.}, $pn\rightarrow
\Lambda\Theta^+$, $pp\rightarrow \Sigma^+  \Theta^+$, and
$pp\rightarrow \pi^+ \Lambda \Theta^+$.  Nam {\em et
  al.}~\cite{Nam:2004qy,Nam:2004hk} found the cross sections for
the $NN\rightarrow \Theta Y$ with the positive-parity $\Theta^+$ to be
approximately ten times larger than those with the negative-parity
one.  A similar tendency was found in the photoproduction of the
$\Theta^+$~\cite{Nakayama:2003by,Nam:2003uf,Nam:2004ph}.  Hence, it is
of great importance to understand how the mass and the width of the
$\Theta^+$ can be changed in a medium when its two different parities
are considered.

There are many indications that the fundamental properties of
hadrons, e.g., masses of hadrons, are modified in a medium.  Dilepton
experiments at Gesellschaft f\"ur Schwerionenforschung (GSI) 
showed a clear signature of dropping $\rho$
masses~\cite{Li95} even though some
controversial arguments still remain~\cite{Rapp}. Recent experiments in
relativistic heavy-ion collisions give stronger support for the
dropping masses of hadrons~\cite{shuryakbrown,prakash}.  The mass
shifts of hadrons also play an essential role in understanding
neutron stars~\cite{LLB97,Glendenning:jr,Savage:1995kv}, e.g., the
masses and the sizes of neutron stars, and the cooling history of young
neutron stars~\cite{Pons}.

Recent studies of the $\Theta^+$ in the context
of heavy-ion collisions suggest that the $\Theta^+$ yield may provide
information on the early stage of a heavy-ion collision due to its weak
interaction with other hadrons produced in the course of the
collision~\cite{heavyion}.  In this regard, it is of great importance
to understand how the $\Theta^+$ is modified in dense matter.

Our aim in this work is to investigate medium modifications of
the $\Theta^+$ baryon, focusing on the parity of the $\Theta^+$.  We
will demonstrate how changes of its chemical potential in dense
matter depend on its parity. We will also investigate the effect
of scaled nucleon masses in dense medium.

\def\p{p\!\!\!/}
\def\k{k\!\!\!/}

\begin{figure}
\centerline{\includegraphics[height=1.0in]{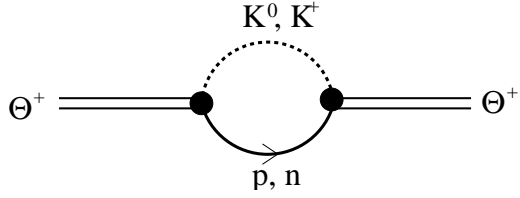}}
\caption{Density dependent self energy contribution to
the $\Theta^+$ propagator. The baryon line $p,n$ indicates
the density dependent propagator of baryons in symmetric
nuclear matter.}
\label{figa}
\end{figure}


We begin by introducing the effective Lagrangian for
$n\Theta K$ coupling~\cite{Nam:2003uf}:
\be {\cal L}_{n\Theta
K} =-\frac{g_A^\star}{2 f_\pi} \bar\Theta \gamma_\mu\gamma_5\left(
\partial^\mu K^+ n -\partial^\mu K^0 p\right) + {\rm h.c.},
\ee
where $g_A^\star$ denotes the pseudo-vector coupling constant for the
$n\Theta K$ vertex.  Here, positive parity of the $\Theta^+$ is
assumed.  Since the parity of the $\Theta^+$ has not been determined yet
from experiments, we also need to consider the Lagrangian for a 
negative-parity $\Theta^+$ in which there is no $i \gamma_5$ in the vertex.  
For numerical estimates, we used $g_A^\star=0.28$ for positive parity
and $g_A^\star=0.16$ for negative parity. These coupling constants
are fixed to reproduce the decay width $\Gamma_{\Theta^+}=15$ MeV.
The
nucleon propagator in a dense medium with the Fermi momentum $k_F$ can be
written as~\cite{GBR98}
\be
G^0_{\alpha\beta} &=& (\gamma_\mu k^\mu +M)_{\alpha\beta}
\left[\frac{1}{k_\lambda^2-M^2+i\epsilon} +\frac{i\pi}{E^0 (k)}
\delta \left(k^0-E^0(k)\right) \theta(k_F-|\vec k|)\right]
\nonumber\\
&=& G_F^0 (k)_{\alpha\beta} + G_D^0 (k)_{\alpha\beta},
\ee
where $E^0(k)=\sqrt{{\vec k}^2+M^2}$, $G_F^0(k)$ indicates the nucleon
propagator in free space and $G_D^0 (k)$ indicates the density-dependent
part of the propagator. For simplicity, we consider symmetric
nuclear matter.  Note that the density is given as
\be
\rho_N=\frac{\gamma}{6\pi^2} k_F^3
\ee
with a degeneracy $\gamma=4$ for symmetric nuclear matter.
We assume that the kaon propagator in dense
matter becomes
\be
\Delta_K^0 (k) &=& \frac{1}{k_\mu^2- m_K^{2} +i\epsilon},
\ee
where $m_K$ is the kaon mass in free space.
There are many experimental and theoretical indications
that the effective masses of baryons and mesons change in a medium.
To be more consistent, one may have to
consider the density-dependent nucleon and kaon properties.
However, since we cannot treat them self-consistently in our analysis
and since our aim is to understand how the properties of $\Theta^+$ change
in a medium, we take the free-space nucleon and kaon masses as
a first-order approximation.  
Later, we consider the density-dependent nucleon mass 
as a second approximation.

The density-dependent part of the $\Theta$ self-energy can be
obtained by considering the contribution of $G_D^0 (k)$
to the diagram in Fig.~\ref{figa}:
\be
\Sigma(p) &=& \frac{1}{i}\frac{i^4}{(2\pi)^4}
     \left|\frac{g_A^\star}{2 f_\pi}\right|^2
\int d^4 k \left[ (p\!\!\!/ -k\!\!\!/) \gamma_5 \Delta_K^0 (p-k)
            G_D^0 (k) \gamma_5  (p\!\!\!/ -k\!\!\!/)  \right]
            \nonumber\\
&=& \Sigma^s (p) - \gamma^0 \Sigma^0 (p)
   + \vec \gamma\cdot \vec p \ \Sigma^v (p),
\ee
where $p$ is the incident momentum of the $\Theta$, and $k$
stands for the loop momentum carried by the nucleon in the medium.
%
The $\Theta^+$ propagator in the medium is modified as
\be
G (p) &=& \frac{1}{ \gamma^0 (p_0+\Sigma^0)
-\vec\gamma\cdot\vec p (1+\Sigma^v)
- (M_\Theta +\Sigma^s) +i\epsilon},
\ee
where $M_\Theta$ is the $\Theta^+$ mass in free space.  The chemical
potential $p_0$ of the $\Theta^+$ can be obtained by solving the
equation
\be
\left(p_0+\Sigma^0 (p)\right)^2 - |\vec p|^2 (1+\Sigma^v)^2
- \left(M_\Theta+\Sigma^s(p)\right)^2 =0.
\ee
In the case of the negative-parity $\Theta^+$, we obtain a similar
expression with one opposite sign in $\Sigma^s(p)$
due to the absence of $i \gamma_5$ in the Lagrangian.

In order to estimate the chemical potential in the rest frame
of $\Theta^+$, we consider $\vec p=0$.  There are both real and
imaginary parts in the density-dependent self-energy.  However, since
the imaginary part is very small compared to the
$\Theta^+$ mass in the medium, one can treat the real and the imaginary parts
separately in the first approximation.  Then, the chemical potential
can be obtained from the linear equation
\be
p^\pm_0 = M_\Theta + {\cal R}e \big(\pm \Sigma^s(p_0)-\Sigma^0 (p_0) \big),
\label{eq1}
\ee
where the superscript $\pm$ indicates the parity of the $\Theta^+$ in the
Lagrangian. 
The contribution to the decay width can also be obtained in this order
by 
\be
{\Gamma^\pm (p_0)} = 2\times
{\cal I}m \big(\pm \Sigma^s(p_0) - \Sigma^0 (p_0) \big),
\label{eq2}
\ee
where $p_0$ is the solution of Eq.~(\ref{eq1}).
However, since the phase space for $\Theta^+ \rightarrow KN$ disappears
beyond $k_F\approx 260$ MeV, detailed investigations of the decay width
are not required. Thus, we will concentrate on the effective masses 
in this work.

\begin{figure}
\centerline{\includegraphics[height=4in]{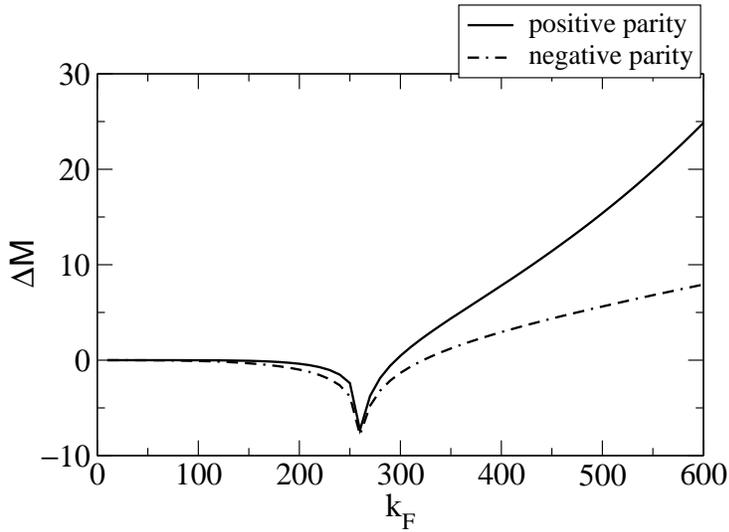}}
\caption{Shifts in the chemical potential of $\Theta^+$ in
symmetric nuclear matter.
The solid curve is for positive parity $\Theta^+$
whereas the dot-dashed curve is for negative parity $\Theta^+$.
}
\label{figb}
\end{figure}

In Fig.~\ref{figb}, the shifts $\Delta M$ in the chemical potential of
$\Theta^+$ in a medium are summarized in the solid curve for the
positive-parity $\Theta^+$ and the dot-dashed curve for the negative-parity
one.  
In some $k_F$ region, there is more than factor of two  difference
between the positive and the negative parity states.


\begin{figure}
\centerline{\includegraphics[height=4in]{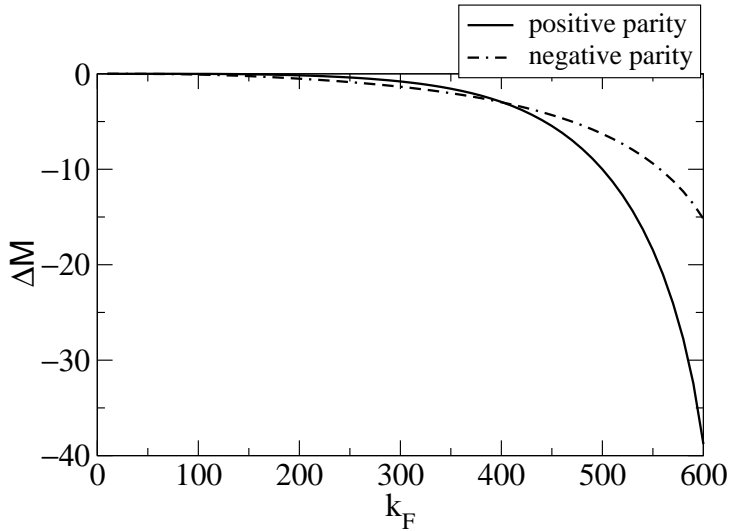}}
\caption{Shift in the chemical potential of the $\Theta^+$
in symmetric nuclear matter
with scaled nucleon masses. When the reduced nucleon
effective mass is applied, the kink disappear, 
and the shift shows a monotonic behavior.
The notations are the same as in Fig.~\ref{figb}}
\label{figd}
\end{figure}

In the previous section, we obtained the density-dependent
results in the first approximation where there are no shifts
in the nucleon and the kaon masses.  However, the
effective masses of nucleons and kaons
are well known to change in a medium.  In order to investigate the effects of
these density-dependent masses, one has to solve the equation
of state for the nucleon more consistently.  However, since our goal
is to see the properties of the $\Theta^+$ not the nucleon itself, we
test the possible effects of the density-dependent nucleon mass by
considering its simple scaling in Eqs.~(\ref{eq1}) and (\ref{eq2}):
\be
M_N^\star = \frac{1}{1+0.23 \rho/\rho_0} M_N.
\ee
The scaling for the kaon is rather uncertain, so 
we have assumed the kaon mass to be constant in our estimate drawn in
Fig.~\ref{figd}.  The kinks in Fig.~\ref{figb} disappear
when the nucleon mass is scaled, and the shift of the $\Theta^+$ mass
turns out to be monotonic.  Note that the parity inverts the sign in the
mass shifts as $k_F$ increases because the pole
(when $M_\Theta = E_N +E_K$) disappears due to the reduced nucleon mass.  

%
In this letter, we investigated the medium modification of
the $\Theta^+$ chemical potential. 
We also investigated the effects of the density-dependent nucleon
masses.  We found that the contributions to the chemical potential  
were very sensitive to the scaled nucleon mass and the parity of the
$\Theta^+$.  While we considered the $KN$ one-loop diagrams only,
possible contributions of particle-hole diagrams, such as $Yh$ and
$Y^*h$, may not be negligible in describing the medium modification of
the $\Theta^+$~\cite{Cabrera}.  

Most of the $\Theta^+$ produced in a dense medium will decay
outside of the fireball due to the very narrow decay width.
Thus, the change in the chemical potential may not be easily detected
in experiments. Furthermore, current experiments cannot 
provide good enough statistics for the $\Theta^+$ production,
at the moment; thus, it may be hard to obtain meaningful information on the 
parity from $\Theta^+$ production in heavy ion collisions.  
However, if the momentum of the $\Theta^+$ is small enough, 
its medium modification may be detected.  For example, 
the $\Theta^+$ produced via $\pi + N\rightarrow \Theta^+ +
\overline{K}$ near threshold can have a small momentum~\cite{ahn}.  In
that case, the $\Theta^+$ may be bound inside the nuclei.  A recent 
study~\cite{Cabrera} supports a possible bound of the $\Theta^+$
inside light and medium nuclei.  Hence, we believe that the present
investigation will provide a guideline for future experiments and
for related theories.

\vskip 5mm


We would like to thank H. Kim (Yonsei Univ., Korea),
J. K. Ahn (Pusan Nat'l Univ., Korea \& Laser Electron Photon Experiment at
SPring-8) 
and I. K. Yoo (Pusan Nat'l Univ., Korea \& NA49 Collaboration at CERN)
for useful comments and discussions.  HChK would like to
express his gratitude to A. Hosaka and S. I. Nam for valuable
discussions.  The work of HChK is supported by a Korea Research
Foundation grant (KRF-2003-041-C20067).  The work of CHL and HJL is
supported by a Korea Research Foundation grant (KRF-2002-070-C00027).

\end{document}